\documentclass[twocolumn,aps,prb]{revtex4-1}
\usepackage{graphicx}
\usepackage{amssymb}
\usepackage{amsmath}
\usepackage{bm}
\usepackage{color}

\def\Journal #1,#2,#3,#4#5#6#7{#1 {\bf #2}, #3 (#4#5#6#7)}

\def\Vec{\mathbf}
\def\GVec#1{\mbox{\boldmath $#1$}}

\def\lsim{\, \lower -0.3ex \hbox{$<$} \kern -0.75em \lower 0.7ex \hbox{$\sim$} \,}
\def\gsim{\, \lower -0.3ex \hbox{$>$} \kern -0.75em \lower 0.7ex \hbox{$\sim$} \,}

\begin{document}

\title{Quantum Hall effect in three-dimensional graphene \\
}
\author{Toshiki Kiryu and Mikito Koshino}
\affiliation{Department of Physics, Osaka University,  Toyonaka 560-0043, Japan}
\date{\today}

\begin{abstract}
We theoretically study the electronic band structure and the Hall effect in the negatively-curved three-dimensional (3D) graphene network in magnetic fields.
We found that special energy regions appear above and below the zero-energy Landau level,
where the 3D Hall conductivity is quantized even though the spectrum is not fully gapped.
The energy regions are dominated by the chiral snake states traveling along the zero magnetic field contour on the 3D surface,
and the quantization of the Hall conductivity crucially depends on the topology of the contour.
The exclusive energy region for the chiral states is found to be a universal feature of 3D graphene systems.
\end{abstract}

\maketitle


\section{Introduction}

\begin{figure}
\begin{center}
\leavevmode\includegraphics[width=0.95\hsize]{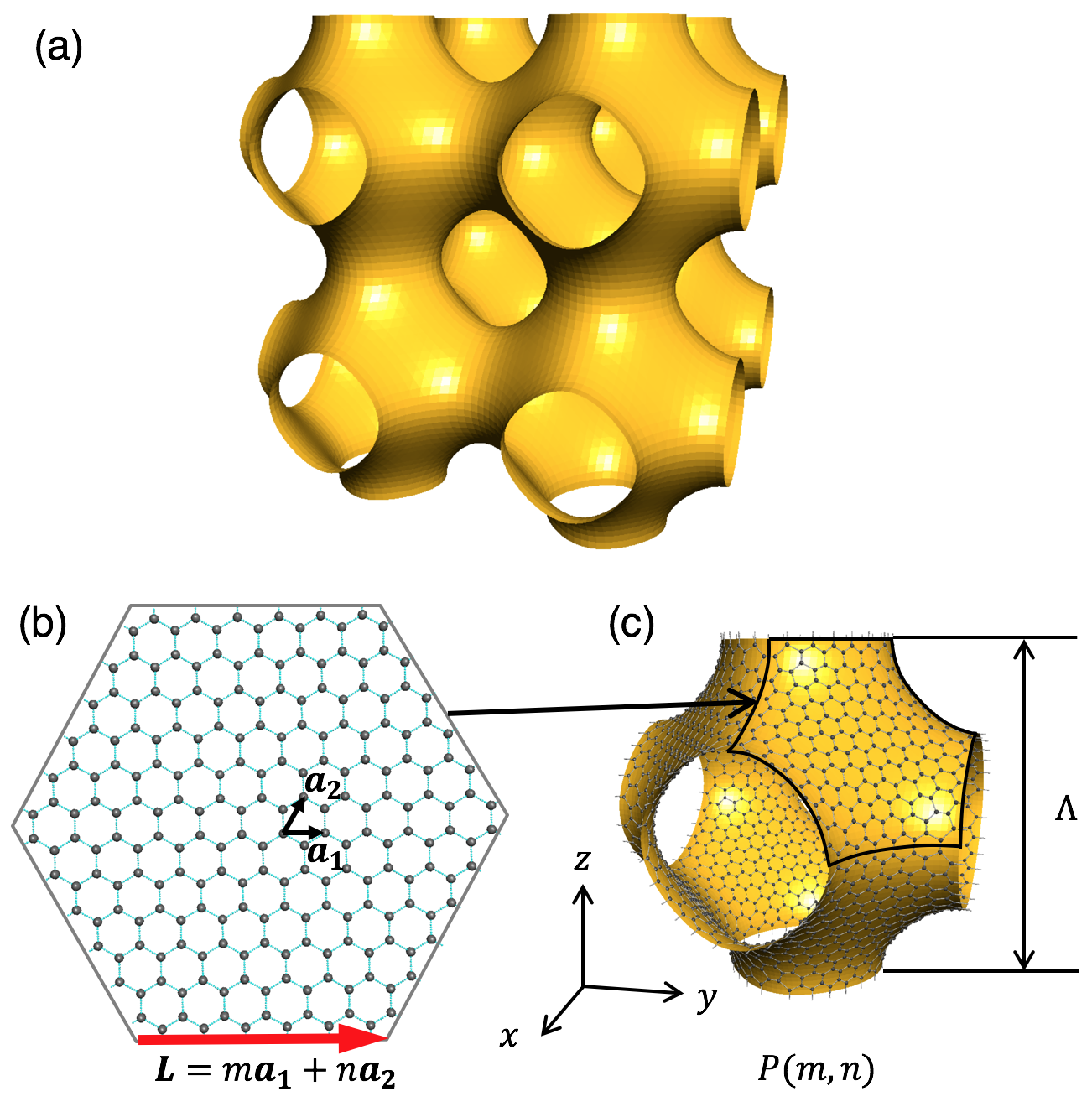}
\end{center}
\caption{(a) Structure of a 3D periodic surface considered in this work (P-surface).
(b) Primitive graphene patch with a hexagonal shape characterized by $\Vec{L}$. 
Here $\Vec{a}_1 , \Vec{a}_2$ are primitive vectors for the honeycomb lattice.
(c) 3D graphene network obtained by fitting the primitive patch to the P-surface.
}
\label{fig_patch}
\end{figure}

Graphene is a versatile material which allows many different geometrical configurations.
Not only its intrinsic two-dimensional (2D) form \cite{novoselov2004electric} 
and low-dimensional forms such as fullerene \cite{kroto1985c60} and carbon nanotube \cite{iijima1991helical}, 
it is theoretically possible to have a three dimensional (3D) curved surface network extending over the space as in Fig.\ \ref{fig_patch},
where the topological disclinations \cite{gonzalez1992continuum,PhysRevB.49.7697,lammert2004graphene}
(e.g. seven-membered ring) generates the negative Gaussian curvature.
\cite{mackay1991diamond,terrones1992geometry,vanderbilt1992negative,lenosky1992energetics,fujita1995polymorphism, weng2015topological, tagami2014negatively, koshino2016dirac} 
Experimentally, the 3D carbon nanostructures were fabricated using various methods,\cite{ma2001very,ma2002synthesis,nishihara2009possible,li2013high,wu2012three,xu2013flexible,chen2011self,yang2013liquid,chen2011three,cao2011preparation,jiang2014design,ito2014high,tanabe2016electric,qin2017mechanics}  
and in particular, a recent experiment successfully realized a smooth and high-quality 3D curved surface network composed of monolayer graphene.
 \cite{ito2014high,tanabe2016electric}

The 3D graphene offers a unique platform where the massless Dirac electrons are confined on the 3D labyrinth of the curved surface.
In this paper we ask: how are the Landau levels formed and how does the quantum Hall effect (QHE) 
occur when such a system is subjected to a magnetic field?
Usually, the QHE occurs in 2D electron systems where the quantization of the Hall conductivity 
relies on the existence of the energy gap between the discrete Landau levels.
The 3D graphene can be viewed as intermediate between 2D and 3D,
in that the electrons locally move on 2D surface while the trajectory viewed in the large scale is 3D.
It is non-trivial how the overall electronic spectrum and transport property look like under magnetic field.

The electronic band structures of 3D curved surface systems were theoretically studied
for 3D periodic graphenes \cite{fujita1995polymorphism, weng2015topological, tagami2014negatively, koshino2016dirac}, 
as well as conventional non-relativistic electrons \cite{aoki2001electronic,aoki2004electronic,koshino2005electronic}  
Recently one of the authors studied the energy spectrum of 3D periodic graphenes,
and showed that an electron behaves as a massive Dirac electron with a scalable mass, 
which inversely scales to the super structure period. \cite{koshino2016dirac}
However, the electronic properties of 3D graphene under magnetic field  are yet to be investigated.

Here we calculate the electronic structure and the Hall conductivity in periodic 3D graphene systems in magnetic fields.
We found that special energy regions appear above and below the zero-energy Landau level,
where the QHE can occur even though the spectrum is not fully gapped.
The energy region is dominated by the chiral states traveling along the zero magnetic field contour,
and the quantization of the Hall conductivity crucially depends on the topology of the chiral-state trajectory, which is determined by the magnetic field orientation
and the topology of the 3D surface.

The paper is organized as follows. We first introduce the theoretical methods in Sec.\ \ref{sec_theor},
and present the band structure and wave functions in Sec.\ \ref{sec_band}. 
In Sec.\ \ref{sec_hall}, we calculate the Hall conductivity and argue about the relationship between
the QHE and the spatial connection of the chiral-state trajectory.
In Sec.\ \ref{sec_disc},  we consider a carbon nanotube in a strong magnetic field to explain the origin of 
the exclusive energy window dominated by the chiral states.
A brief conclusion is given in Sec.\ \ref{sec_concl}.

\begin{figure*}
\begin{center}
\leavevmode\includegraphics[width=0.8\hsize]{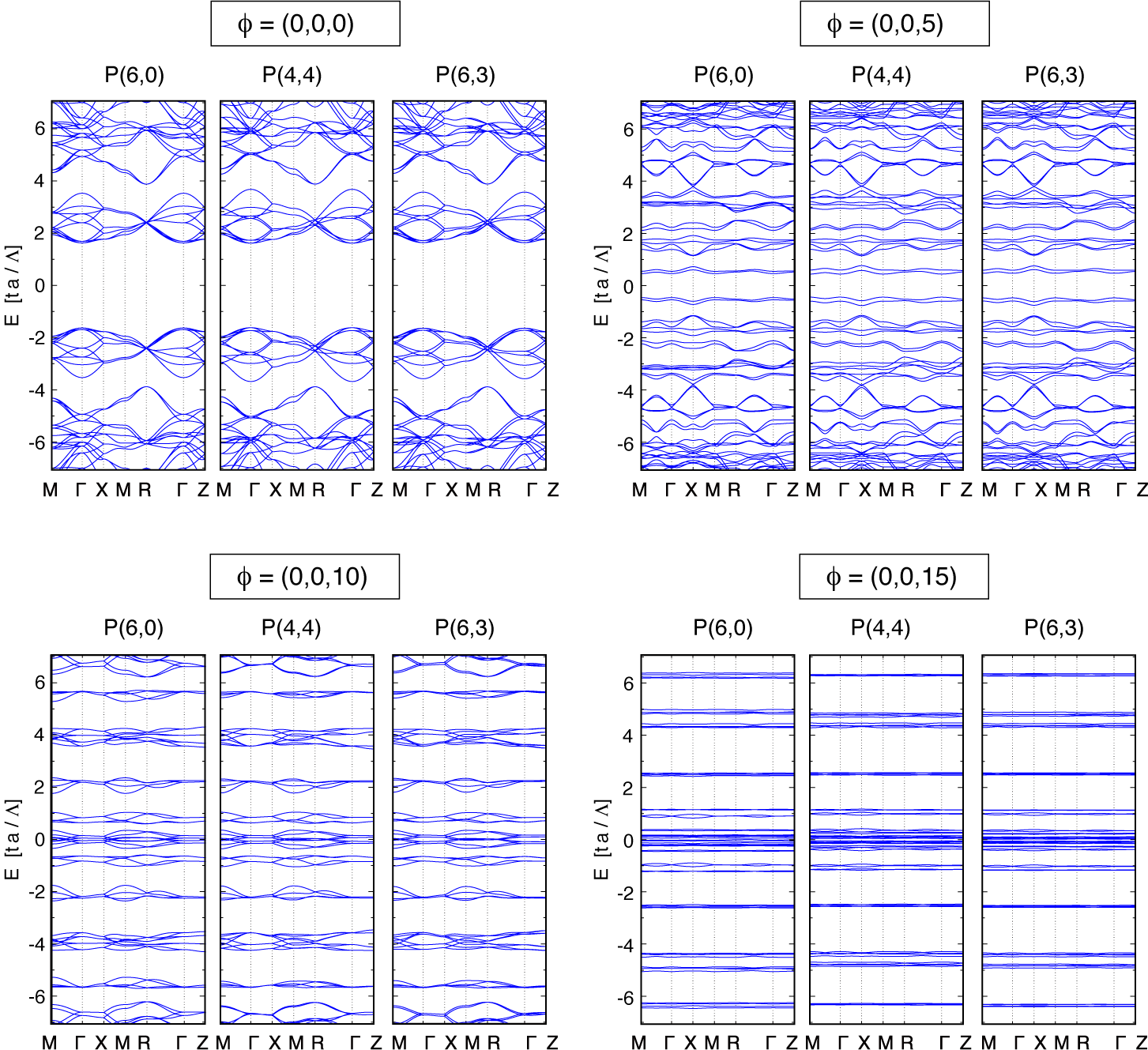}
\end{center}
\caption{Band structures of 3D graphene systems $P(6,0), P(4,4), P(6,3)$
at magnetic fluxes $\phi=(0,0,0), (0,0,5), (0,0,10)$ and (0,0,15).}
\label{fig_family}
\end{figure*}

\section{Theoretical methods}
\label{sec_theor}

We consider the P-surface as a periodic network of continuous membrane depicted in Fig.\ \ref{fig_patch}(a).
The graphene network having the P surface topology
is constructed by taking the primitive hexagonal patch of Fig.\ \ref{fig_patch}(b), and deforming it to fit the curved surface as in Fig.\ \ref{fig_patch}(c).
The primitive patch is characterized by indeces $(m,n)$, which specify 
the lattice vector connecting the neighboring corners of the hexagon,
\begin{align}
{\bf L} = m {\bf a}_1 + n {\bf a}_2,
\label{eq_L}
\end{align}
where ${\bf a}_1=a(1,0)$ and ${\bf a}_2=a(1/2,\sqrt{3}/2)$ are primitive unit vectors
of the original graphene honeycomb lattice.  
The resulting 3D structure is denoted by $P(m,n)$. 
Its cubic unit cell [Fig.\ref{fig_patch} (c)] is composed of eight primitive patches,
and it is periodic with 3D lattice vectors $\Vec{\Lambda}_j = \Lambda \, \Vec{e}_j \, (j=x,y,z)$,
where $\Lambda = 2\sqrt{2} |\Vec{L}| = 2[2(m^2+n^2+mn)]^{1/2} a$ and $\Vec{e}_j$ is the unit vector along $j$ direction.
The atomic structure has an eight-membered ring at each vertex where four patches meet,
while it is composed of six-membered rings elsewhere. 
The deformation of the hexagonal patch onto the curved surface is performed as the follows.
We project the curved patch [enclosed by black line in Fig.\ \ref{fig_patch}(c)]  to $(1,1,1)$ direction,
and then deform the flat object to a hexagon with the six vertexes fixed,
by radially inflating it with the constant ratio depending on the direction.
In this way, we obtain one-to-one mapping between the flat hexagon and the curved surface.

We model the system with the simple tight-binding Hamiltonian for $p_z$ orbital of carbon,  
\begin{align}
& H= -t \sum_{\langle i,j \rangle} e^{i\theta_{ij}}c^\dagger_i c_j + {\rm h.c.},  \label{eq_H}\\
& \theta_{ij} = -\frac{e}{\hbar}\int_{\Vec{R}_j}^{\Vec{R}_i} 
\Vec{A}(\Vec{r})\cdot d\Vec{r}, \label{eq_peierls} 
\end{align}
where $c^\dagger_i$ and $c_i$ are creation and annihilation operators of an electron on $i$-th site,
$t$ is the hopping integral, $\langle i,j \rangle$ labels pairs of the nearest-neighbor sites
and $\Vec{R}_i$ is the 3D position of the $i$-th site.
Here we neglect the curvature effect on the hopping parameter $t$, and 
take it as a constant ($t\simeq 3$ eV) regardless of the position.
Note that $\theta_{ij}$ is the Peierls phase induced by the external magnetic field,
where $\Vec{A}(\Vec{r})$ is  the vector potential which gives a uniform magnetic field
$\Vec{B}=\nabla \times \Vec{A}$.
We adopt the gauge $\Vec{A}(\Vec{r}) = (B_y z, B_z x, B_x y)$ in the following.
The integral in Eq.\ (\ref{eq_peierls}) is performed on a straight line from $\Vec{R}_j$ to $\Vec{R}_i$.

\begin{figure*} 
\begin{center}
\leavevmode\includegraphics[width=0.8\hsize]{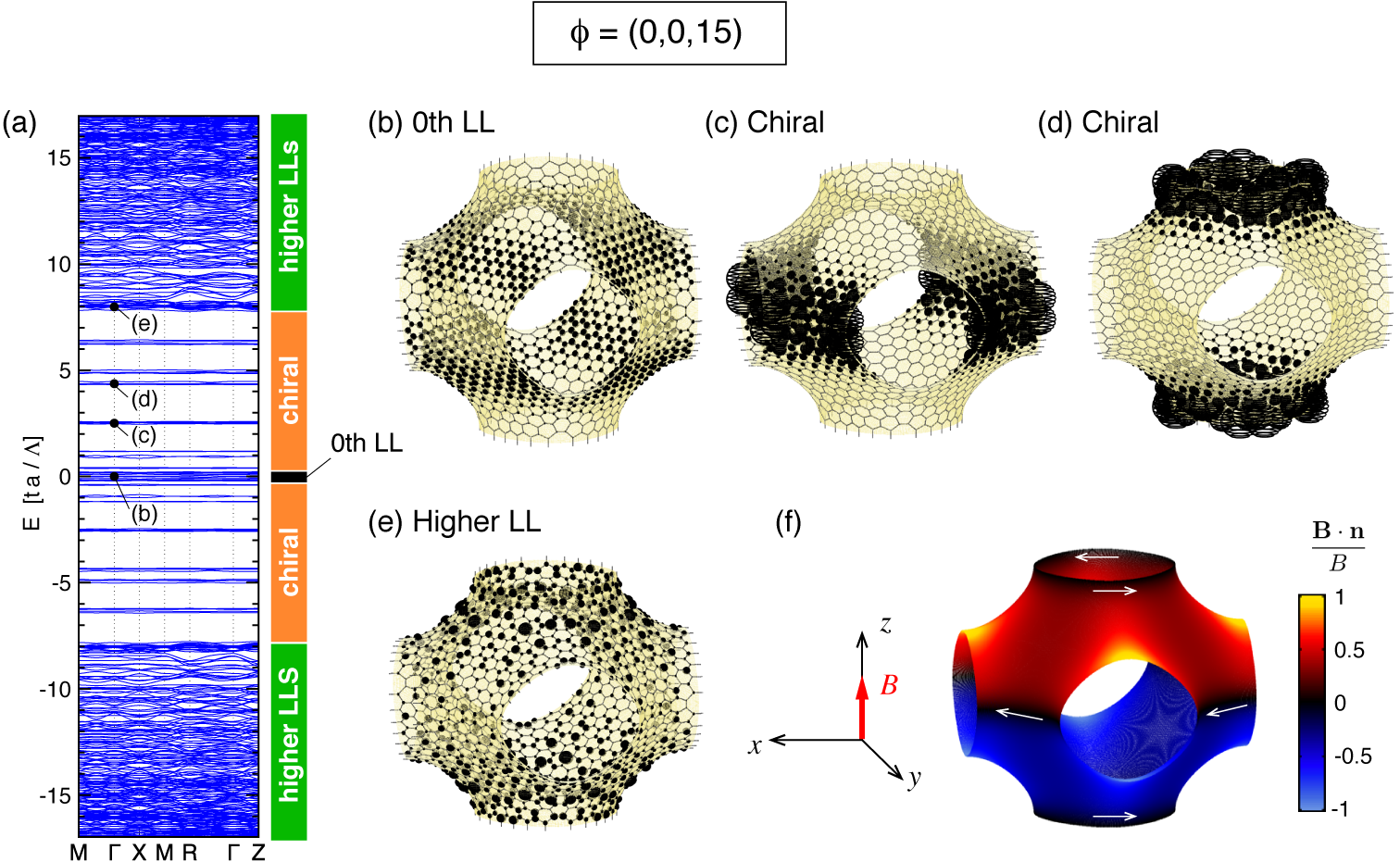}
\end{center}
\caption{
 (a) Energy spectrum of $P(6,0)$ at $\GVec{\phi}=(0,0,15)$.
 (b), (c), (d), (e) Wave functions of representative states of which eigen energy is indicated in (a).
 (f) Map of $B_\perp =0$, or the magnetic field component normal to the curved surface.
}
\label{fig_001}
\end{figure*}

\begin{figure}
\begin{center}
\leavevmode\includegraphics[width=0.8\hsize]{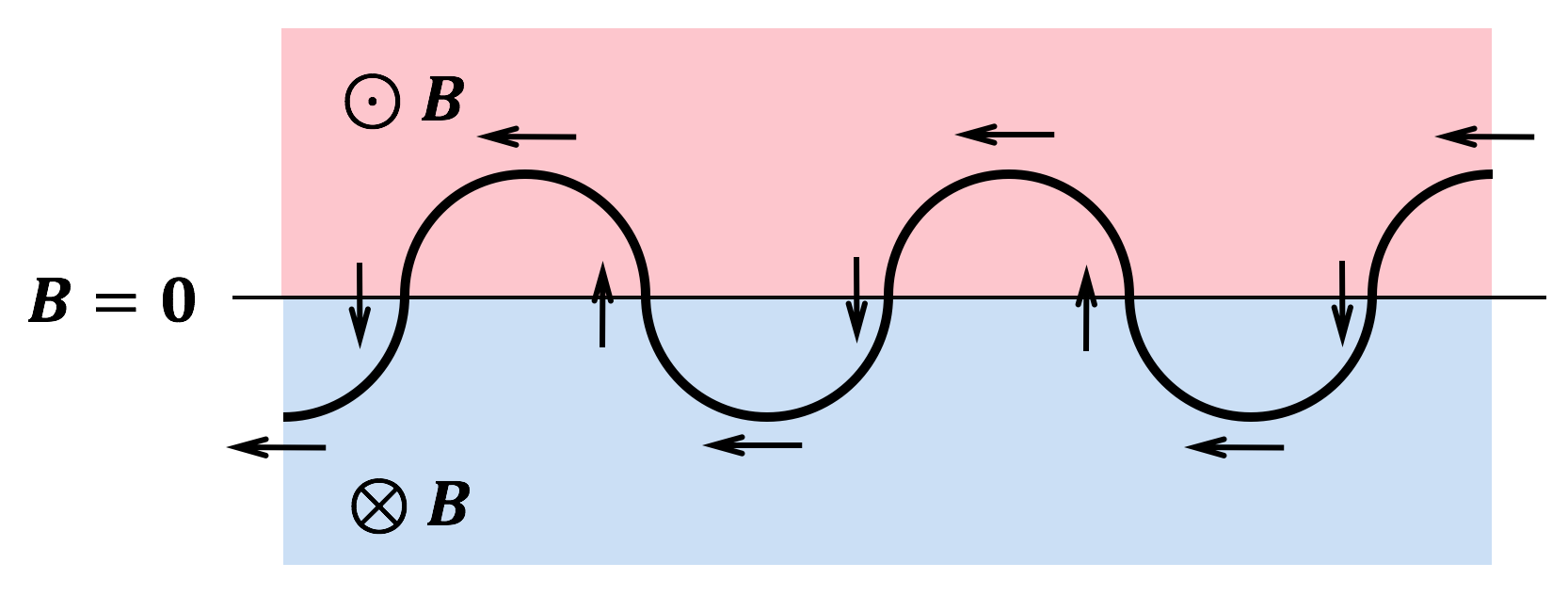}
\end{center}
\caption{Classical picture of chiral snake orbits emerging between the positive and negative magnetic field regions.}
\label{fig_snake}
\end{figure}

\begin{figure*}
\begin{center}
\leavevmode\includegraphics[width=0.8\hsize]{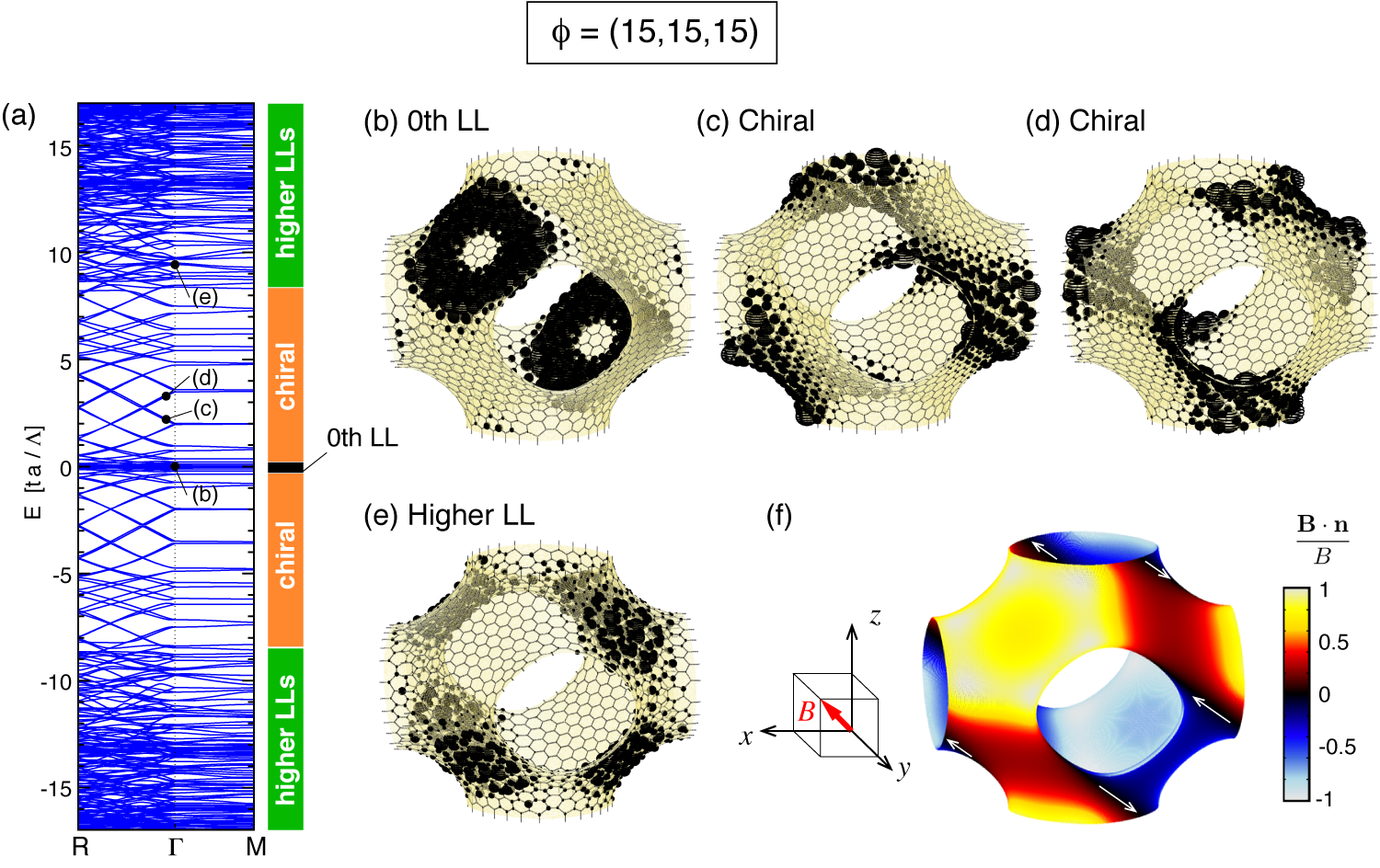}
\end{center}
\caption{Set of plots similar to Fig.\ \ref{fig_001}, obtained for $\GVec{\phi}=(15,15,15)$.
}
\label{fig_111}
\end{figure*}

The system is characterized by
the number of magnetic flux quanta penetrating the unit cell faces,
\begin{align}
\GVec{\phi} \equiv (\phi_x,\phi_y,\phi_z) = (B_x,B_y,B_z) \frac{\Lambda^2}{h/e}.
\end{align}
In magnetic field, the Hamiltonian is no longer translationally-symmetric
because of the spatial dependence of the vector potential.
When $\phi_j$'s are rational numbers, however, 
we can introduce a magnetic unit cell and construct the eigenstates
so as to satisfy the magnetic Bloch condition.\cite{brown1968ascpects,RevModPhys.82.1959}
When $\phi_x$, $\phi_y$ and $\phi_z$ are all integers, in particular,
the magnetic unit cell becomes just the same as the original unit cell spanned by $\Vec{\Lambda}_j$'s,
where the magnetic Bloch condition becomes
\begin{eqnarray}
 \Psi_\Vec{k}(\Vec{r}+\Vec{\Lambda}_j) &=& 
e^{i \Vec{k}\cdot \Vec{\Lambda}_j}
e^{-i (e/\hbar)(\Vec{A}- \Vec{B}\times \Vec{r})\cdot\Vec{\Lambda}_j}
 \Psi_\Vec{k}(\Vec{r}) 
 \label{eq_mag_bloch}
\end{eqnarray}
for $j=x,y,z$ directions, where $\Psi_\Vec{k}(\Vec{r})$ is the wave amplitude of the site at $\Vec{r}$,
and $\Vec{k}$ is the Bloch wavenumber defined in the 3D Brillouin zone.
Using the condition Eq.\ (\ref{eq_mag_bloch}), we construct the Hamiltonian matrix of Eq.\ (\ref{eq_H})
within a single unit cell, and obtain the energy spectrum by diagonalizing it.

The Hall conductivity is calculated by the linear response formula,
\begin{align}
& \sigma_{\mu \nu} =
g_s \frac{\hbar e^{2}}{V} 
\sum_{n\Vec{k}} 
 f(E_{n\bm{k}})
\nonumber\\
& \quad \times
\sum_{m (\neq n)}  
2 {\rm Im} \frac{
 \langle n, \bm{k}  | v_\mu |  m, \bm{k} \rangle
   \langle m, \bm{k}   | v_\nu |   n, \bm{k} \rangle
      }
{(E_{n\bm{k}} - E_{m\bm{k}})^{2}},
\end{align}
with the velocity operator,
\begin{align}
& v_\mu =  \frac{1}{\hbar} \frac{\partial H(\bm{k})}{\partial k_{\mu}}.
\end{align}
Here $g_s=2$ is the spin degeneracy, $V$ is the total system volume, $f(\varepsilon)$ is the Fermi distribution function, 
$H(\bm{k})$ is the Bloch Hamiltonian of the tight-binding system, and
$|n, \bm{k} \rangle$ and $E_{n\bm{k}}$ are the $n$-th eigenstate and eigenenergy of $H(\bm{k})$, respectively.
In the present study, the summation in $\Vec{k}$ is taken over discrete $6 \times 6 \times 6$ points in the Brillouin zone.

\section{Band structures in magnetic fields}
\label{sec_band}

Figure \ref{fig_family} shows the energy spectrum calculated 
for 3D graphenes of different unit cell sizes, $P(6,0)$, $P(4,4)$ and $P(6,3)$,
in magnetic fluxes $\GVec{\phi}=(0,0,0)$, $(0,0,5)$, $(0,0,10)$  and $(0,0,15)$.
Here the energy is renormalized in units of $ta/\Lambda$, which depends on the length scale of the 3D structure.
According to the family effect\cite{koshino2016dirac}, the three systems belong to the same class $m-n=0$ in modulo of 3, 
which means that they are all described by the same continuum model in the renormalized scale. 
Indeed,  Figure \ref{fig_family} shows that they have similar energy spectra in zero magnetic field and also in finite magnetic fields as well.
The characteristic energy scale $ta/\Lambda$ is inversely proportional to the length scale $\Lambda$,
and this reflects that fact that the low-energy spectrum is effectively described by the massless Dirac equation linear in the momentum.
As a universal behavior, we see that the original semiconducting gap in zero magnetic field
gradually shrinks in increasing $\phi$, and the band lines are flattened and shift toward the zero-energy.
The results obtained below can be generalized to 3D graphenes with larger unit cell sizes, just by scaling the energy axis.
As we confirmed the scaling behavior of the low-energy spectrum, we will limit our consideration to $P(6,0)$ in the following.


Figure \ref{fig_001}(a) plots the energy spectrum of $P(6,0)$ at $\GVec{\phi}=(0,0,15)$ in the greater energy scale.
At $E \sim 0$, we have a bunch of nearly flat bands corresponding to the 0th Landau levels of graphene.
The high energy region $E \gsim 8(ta/\Lambda)$ is densely filled with band lines, that are later shown to be higher Landau levels.
In the intermediate region $0 < E \lsim 8(ta/\Lambda)$, we have sparsely distributed flat bands which we call the chiral states.
In Figs.\ \ref{fig_001} (b), (c), (d) and (e), we plot the wave function of representative states from each energy region,
of which eigen energy is indicated in Fig.\ \ref{fig_001}(a).
The panel (f) illustrates the map of  $B_\perp$, or the local magnetic field component normal to the curved surface.
We notice that the wave function of (b) (at zero energy) spreads over the region where $B_\perp$ is relatively strong,
while the wave functions of (c) and (d) (sparse distributed area) are concentrated around the contour of $B_\perp = 0$.
Actually the states (c) and (d)  correspond to classical snake orbits as illustrated in Fig.\ \ref{fig_snake},
which emerge on the boundary separating the positive and negative $B_\perp$ regions. \cite{reijniers2000snake}
There the electrons can travel only along the $B_\perp = 0$ contour in a single direction, 
and in that sense we call them the chiral states.
In the magnetic field along $z$ dirction, all the snake orbits encircle the necks of the pipe structure,
and they are disconnected to each other. 
In quantum mechanics, those zero-dimensional states result in 
the discrete energy levels, which correspond to the flat bands in the chiral state region in Fig.\ \ref{fig_001}(a).

 The band structure of the chiral states sensitively depends on the magnetic field direction.
Figure \ref{fig_111}  shows the same set of plots obtained for $\GVec{\phi}=(15,15,15)$.
Now the $B_\perp = 0$ contour becomes open along  $(1,1,1)$ direction as shown in Fig.\ \ref{fig_001}(f),
and the wave functions (c) and (d) follow the $B_\perp = 0$ contour.
Accordingly, the energy bands in the chiral state region aquire one-dimensional nature,
dispersing in $\Gamma-R$ direction ($(1,1,1)$ direction) while almost flat in the perpendicular direction.

\begin{figure*}
\begin{center}
\leavevmode\includegraphics[width=1.\hsize]{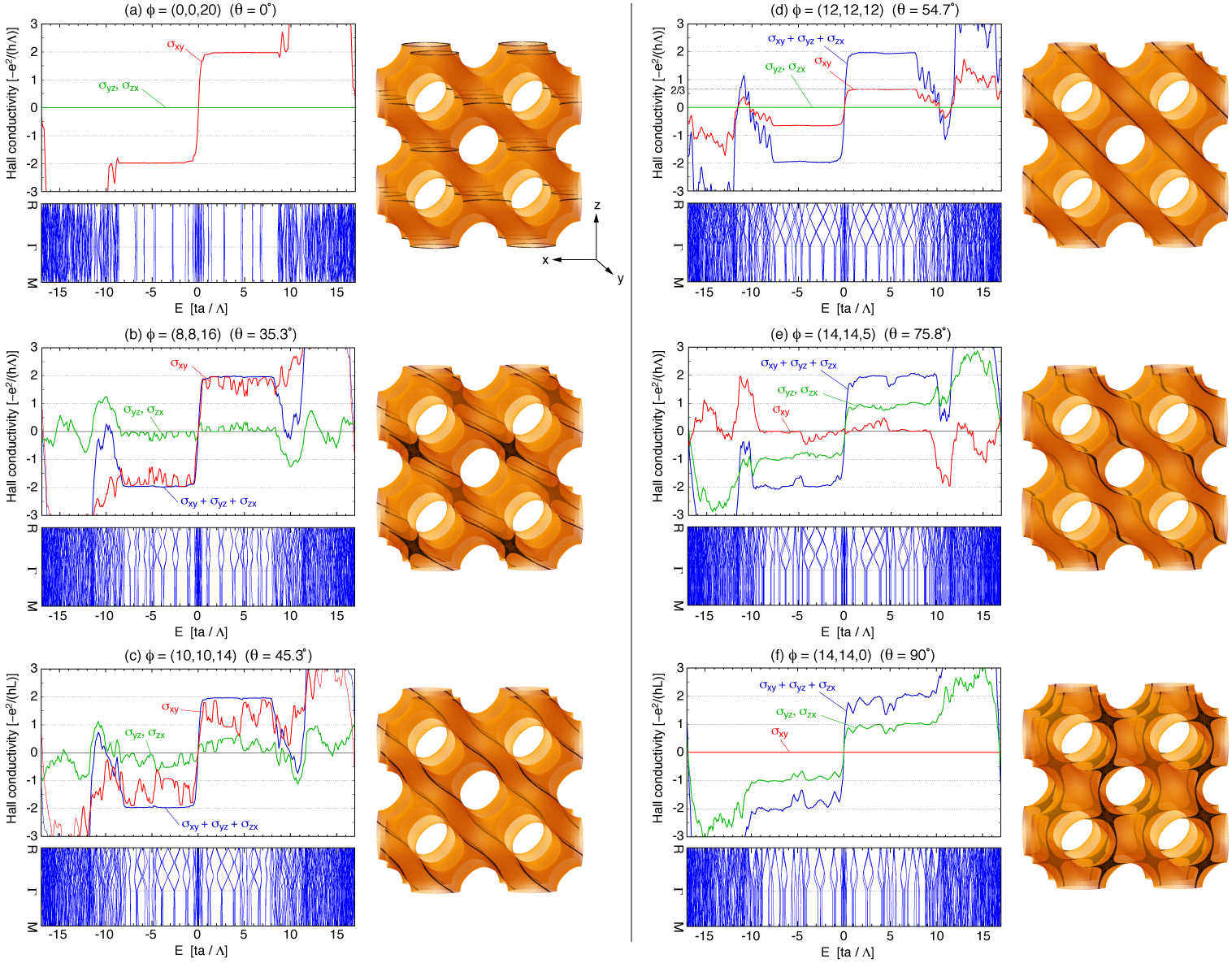}
\end{center}
\caption{Plots of the Hall conductivity tensors, $\sigma_{xy}, \sigma_{yz}$ and $\sigma_{zx}$,
and $\sigma_{xy} +\sigma_{yz} + \sigma_{zx}$
as functions of the Fermi energy, calculated for $P(6,0)$.
The different panels from (a) to (f) are for different magnetic field orientation,
which is tilted from (0,0,1) to (1,1,0) with the angle $\theta$.
In each panel, the energy band structure is shown in the bottom,
and the $B_\perp =0$ contour on the 3D structure is shown on the right.
}
\label{fig_hall}
\end{figure*}

\section{Hall conductivity}
\label{sec_hall}

The connectivity of the $B_\perp = 0$ contour is closely related to the quantization of the Hall conductivity.
Figure \ref{fig_hall} presents the plots of the Hall conductivity tensors, $\sigma_{xy}, \sigma_{yz},\sigma_{zx}$, and $\sigma_{xy} +\sigma_{yz} + \sigma_{zx}$
as functions of the Fermi energy, calculated for $P(6,0)$.
Here the different panels from (a) to (f) are for different magnetic field orientation,
which is tilted from (0,0,1) to (1,1,0) with the angle $\theta$.
The magnitude of $\GVec{\phi}$ is chosen so that the absolute magnetic field is nearly constant in tilting.
In each panel, the energy band structure is shown in the bottom,
and the $B_\perp =0$ contour on the 3D structure is shown on the right.

\begin{figure}
\begin{center}
\leavevmode\includegraphics[width=0.8\hsize]{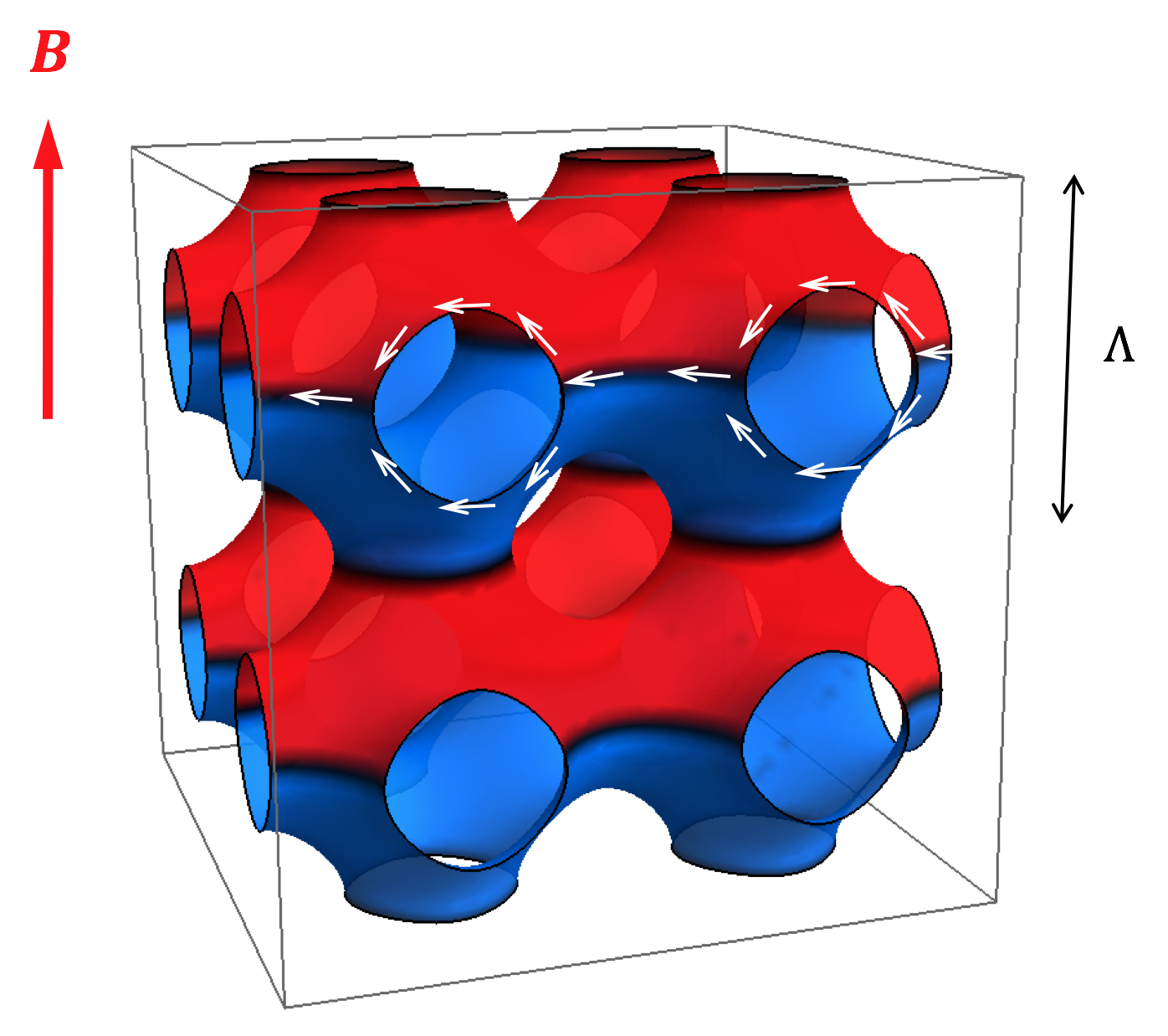}
\end{center}
\caption{
Connection of the edge modes and the $B_\perp=0$ contour modes,
which form surface circulating channels on the boundary box.
The color map indicates $B_\perp$.
}
\label{fig_edge_current}
\end{figure}

In (0,0,1) direction [Fig.\ \ref{fig_hall}(a)], we notice that 
the Hall conductivity $\sigma_{xy}$ is quantized to $\mp 2e^2/(h\Lambda)$ 
in the chiral-state regions in the electron and hole sides, respectively. 
The positive and negative plateaus are connected by a sharp step contributed by the $n=0$ LL.
The Hall conductivity becomes completely flat in the chiral state region, because
the chiral states are all localized into the ring orbits and do not contribute to the electronic transport.
The quantized value $\mp 2e^2/(h\Lambda)$ can by understood by the bulk-edge correspondence. \cite{hatsugai1993chern}
Figure \ref{fig_edge_current} illustrates $2\times 2\times 2$ units of the periodic surface terminated by a bounding box.
In a magnetic field, one-way traveling modes appear on the edge of the curved surface as in the conventional quantum Hall effect.
Those edge modes are connected to the $B_\perp=0$ contour modes,
to form surface circulating channels on the boundary box, as shown by white arrows in Fig.\ \ref{fig_edge_current}.
Including the spin degree of freedom, we have two surface channels per the thickness $\Lambda$ in $z$-direction,
and therefore the Hall conductivity $\sigma_{xy}$ becomes $-2e^2/(h\Lambda)$ in the electron side.
Note that the physical dimension of the Hall conductivity in 3D is different from 2D's, and it has an extra length scale in the denominator.

By tilting the magnetic field from $z$ axis, the topology of the $B_\perp=0$ contour changes and it directly influences the Hall plateau.
The contours change the connectivity at $\theta \sim 35^\circ$ [Fig.\ \ref{fig_hall}(b)], 
and they are reconstructed into open parallel lines along (1,1,1) direction [Fig.\ \ref{fig_hall}(c)].
Accordingly, the Hall conductivity $\sigma_{\mu\nu}$ is not quantized anymore because the orbits are extended,
but we notice that the sum of the components, $\sigma_{111} = \sigma_{xy}+\sigma_{yz}+\sigma_{zx}$, remains quantized to $\mp 2e^2/(h\Lambda)$
in Figs.\ \ref{fig_hall}(a) to (d).
The quantity $\sigma_{111}$ represents the Hall conductivity on the plane normal to (1,1,1) direction;
more precisely, if we take the $XYZ$ coordinate so that $Z$ is parallel to (1,1,1) and $X$ and $Y$ are perpendicular to it,
then $\sigma_{111}$ is equal to $\sigma_{XY}$. The plateau of $\sigma_{XY}$ can be understood
by considering that the $B_\perp=0$ contours are all parallel to $Z$ direction and disconnected in $X$ and $Y$ directions,
where the chiral states do not carry the current along $XY$ plane.
The (1,1,1) direction [Fig.\ \ref{fig_hall}(d)] is special where $\sigma_{xy}, \sigma_{yz}$ and $\sigma_{zx}$ are all quantized into the fractional value
$\mp 2e^2/(3h\Lambda)$ in the plateaus,
and this is because $\sigma_{xy} = \sigma_{yz} =\sigma_{zx}$ is imposed by the symmetry.
When we further tilt the magnetic field to (1,1,0) direction, the $B_\perp=0$ contour lines starts to link together,
eventually forming a 2D network perpendicular to (1,-1,0) direction [Fig.\ \ref{fig_hall}(f)].
As a consequence, the Hall plateau is not precisely flat any more.

\begin{figure}
\begin{center}
\leavevmode\includegraphics[width=0.7\hsize]{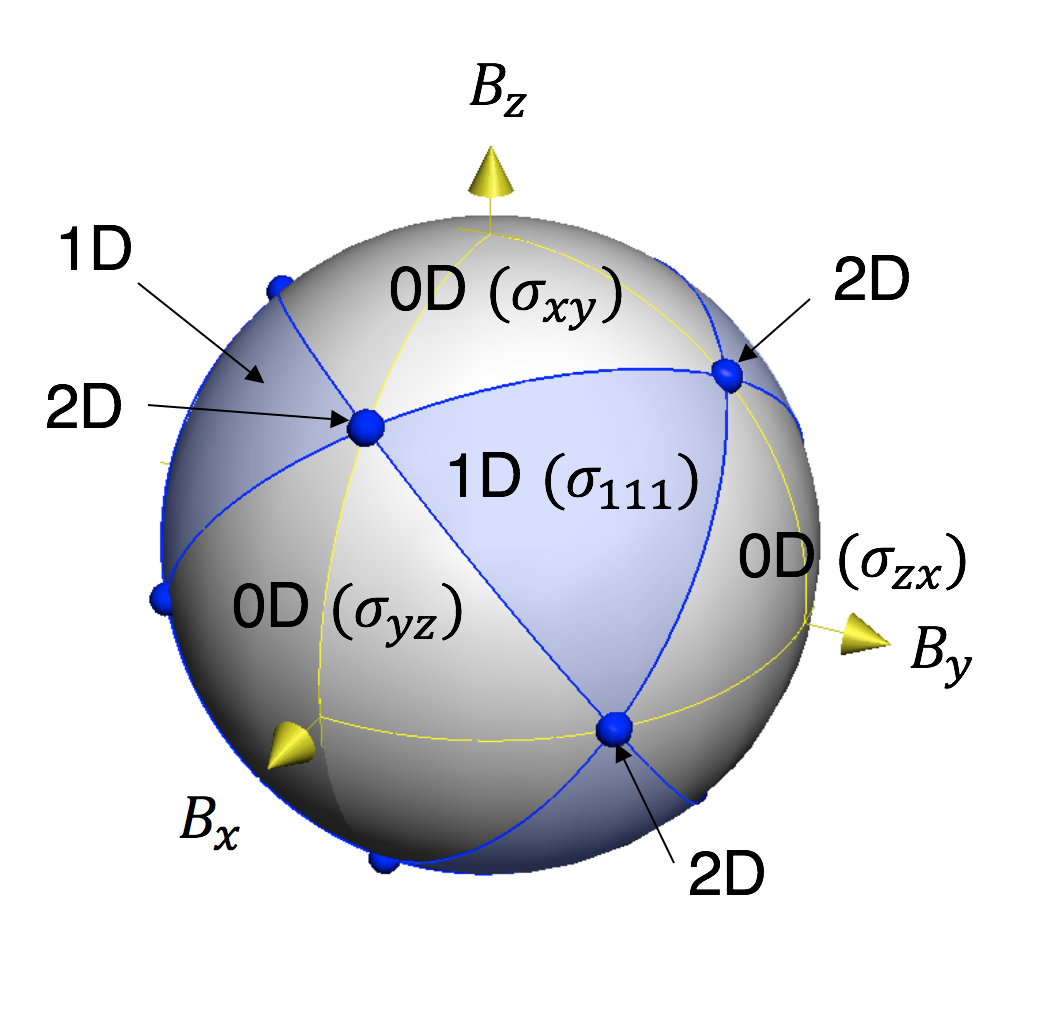}
\end{center}
\caption{Dependence of the connectivity of the $B_\perp=0$ contour 
on the magnetic field orientation. Here 0D, 1D and 2D stand for the dimensionality of the contour network,
and $\sigma_{\mu\nu}$ in the bracket represents the Hall component to be quantized.}
\label{fig_sphere}
\end{figure}

Figure \ref{fig_sphere} summarizes the dependence of the connectivity of the $B_\perp=0$ contour 
on the magnetic field orientation. Here 0D, 1D and 2D stand for the dimensionality of the contour,
and $\sigma_{\mu\nu}$ in the bracket represents the Hall component to be quantized.

\begin{figure}
\begin{center}
\leavevmode\includegraphics[width=1.\hsize]{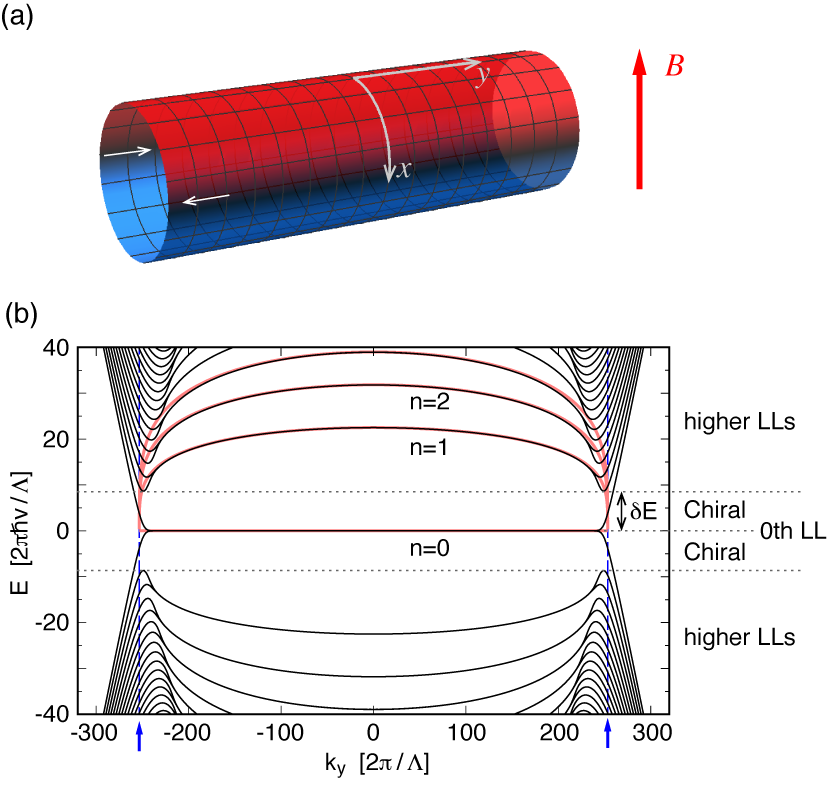}
\end{center}
\caption{(a) Geometry of the carbon nanotube with circumference $\Lambda$ under a magnetic field $B$ perpendicular to the tube axis. 
The color map represents $B_\perp$.
(b) Band structure of the carbon nanotube as a function of $k_y$, in a strong magnetic field of $\Lambda/l_B = 100$. 
Thick pink curves are the Landau levels in the local magnetic field approximation,
and vertical blue arrows indicate the point where the local magnetic field vanishes (see the text).}
\label{fig_nanotube}
\end{figure}

\section{Origin of the chiral-state energy window}
\label{sec_disc}

 The energy window dominated by the chiral states is a characteristic feature
of the curved graphene system in magnetic field, and it is crucial in the emergence of the Hall plateaus.
Naively thinking, it seems that the energy spectrum can be roughly viewed as a collection of the Landau levels of local magnetic field $B_\perp$,
but then it is hard to understand why a finite energy window remains between $n=0$ and $n=1$ Landau levels,
even though $B_\perp$ continuously ranges from zero to the maximum.
Actually, the origin of the energy window can be intuitively understood 
by considering a carbon nanotube in magnetic field. \cite{ajiki1993electronic, ando2005theory}

Let us consider a metallic carbon nanotube with circumference $\Lambda$ under a magnetic field $B$ perpendicular to the tube axis.
We set the two-dimensional coordinate on the tube so that $x$ is parallel to the circumference 
and $y$ is along the tube axis, as in Fig.\ \ref{fig_nanotube}(a).
The magnetic field normal to the surface is $B_\perp(x) = B \cos (2\pi x/\Lambda)$,
and the corresponding vector potential on the surface is  given by $A_x=0$ and $A_y = [B \Lambda / (2\pi)] \sin (2\pi x/\Lambda)$.
The motion of an electron on the tube is described by the 
two-dimensional Dirac Hamiltonian with the periodic boundary condition in $x$, 
and it is expressed as  \cite{ando2005theory,neto2009electronic}
\begin{align}
H =
\begin{pmatrix}
0 & v\pi_- \\
v \pi_+ & 0
\end{pmatrix},
\end{align}
where $\pi_\pm = \xi \pi_x \pm i \pi_y$,
$\pi_i = p_i + eA_i$ and $\xi = \pm$ represents the valley degree of freedom.
Note that $v$ is the graphene's band velocity which is related to the tight-binding hopping integral $t$ by $v = (\sqrt{3}/2) ta/\hbar$.
The eigen function is labelled  by the wave number $k_y$ along the tube axis.
The magnetic field strength is characterized by the dimensionless quantity $\Lambda/l_B$ where $l_B = \sqrt{\hbar/(eB)}$ is the typical magnetic length.

Figure \ref{fig_nanotube}(b) shows the band structure as a function of $k_y$, in a magnetic field of $\Lambda/l_B = 100$.
The most part of the energy spectrum can be regarded as the Landau levels for the local magnetic field $B_\perp(x)$,
which is given by $E_n(B_\perp) = {\rm sgn}(n)\sqrt{2\hbar v^2 e B_\perp n} \, (n=0,\pm 1,\pm 2\cdots)$.
Here the center of the wave function, $x$, is related to $k_y$ by the condition $p_y + eA_y = 0$, or
$\sin (2\pi x/ \Lambda) = - h k_y/ (eB \Lambda)$.
The local field amplitude is then written as a function of $k_y$ as $B_\perp(k_y) = B \sqrt{1-[h k_y/ (eB \Lambda)]^2}$, 
and it vanishes at $k_y=\pm eB\Lambda/h$, which is indicated by the blue vertical arrows.
In Fig.\ \ref{fig_nanotube}(b), the local Landau level energies $E_n(B_\perp(k_y))$ are represented by the thick pink curves,
which agree well with the actual energy levels.

The local Landau level is well defined only when the magnetic field $B_\perp(x)$ is approximately constant inside the wave function.
The spread of wave function is characterized by the local magnetic length $l_B(x) = \sqrt{\hbar /(e B_\perp(x))}$,
and therefore the approximation fails at $B_\perp(x) = 0$ where $l_B(x)$ diverges.
In Fig.\ \ref{fig_nanotube}(b), we actually see that the energy levels significantly deviate from $E_n$ 
near the point of $B_\perp(x) = 0$.
The deviation occurs under the condition of $\delta x \lsim l_B(x)$, where $\delta x$ is the distance from the point of $B_\perp(x) = 0$.
By solving the inequality for $x$,  we obtain
\begin{align}
\delta x \lsim
\left(
\frac{\hbar}{eB}\frac{\Lambda}{2\pi}
\right)^{1/3} \equiv x_0,
\end{align}
which characterizes the spacial region where the local Landau level cannot be well defined.
In Fig.\ \ref{fig_nanotube}(b), the $n=0$ Landau levels split into the positive and negative slopes around this region,
and all other Landau levels $|n| \geq 1$ are also repelled away in a parallel fashion.
As a result, we are left with a finite energy window dominated by the slopes from $n=0$ level, and its energy width is roughly given by
\begin{align}
\delta E  \sim\frac{\hbar v}{x_0} 
= \hbar v \left(
\frac{eB}{\hbar}\frac{2\pi}{\Lambda}
\right)^{1/3}.
\label{eq_delta_E}
\end{align}
This qualitatively explains the energy window of the chiral states in the 3D graphene,
where $\Lambda$ is the unit length of the 3D structure.
By using $\phi = B\Lambda^2/(h/e)$, Eq.\ (\ref{eq_delta_E}) becomes
\begin{align}
\delta E  \sim \frac{\sqrt{3}}{2} \frac{ta}{\Lambda}(4\pi^2 \phi)^{1/3}.
\end{align}
At $\phi = 15$, we have $\delta E \sim 7.3 (ta/\Lambda)$, 
and it agrees with the energy window in $\GVec{\phi}=(0,0,15)$ in Fig.\ \ref{fig_001}(a).

The real 3D graphene sponge fabricated in the experiments\cite{ito2014high,tanabe2016electric} has a randomly connected structure
unlike the periodic model in the present study.
Still, the chiral snake states are expected to emerge on the $B_\perp$ contours in a similar manner,
and its energy width should be characterized by  Eq.\ (\ref{eq_delta_E}), where $\Lambda$ is the typical length scale of the 3D labyrinth.
When we take $\Lambda = 100$ nm \cite{ito2014high,tanabe2016electric},
we have $\delta E \sim 30$ meV and 65 meV for $B=1$ T and 10 T, respectively, which are expected to be within a feasible range for observation.
The emergence of the Hall plateau in the random 3D graphene is a subtle question, because it depends on the localization nature
of the $B_\perp$ contours. The detailed study on the connectivity of $B_\perp$ contours on the random surface
is left for future research.

\section{Conclusion}
\label{sec_concl}

We have calculated the electronic structure and Hall conductivity in the periodic 3D graphenes in magnetic fields.
We found special energy regions right above and below the lowest Landau level,
which are dominated by the chiral snake states traveling along the zero magnetic field contours.
The band structure and the transport property of the chiral states sensitively depend on the topology of the contours,
and we even have a 3D quantum Hall effect when the contours are spatially disconnected.
Finally, we considered a carbon nanotube in a strong magnetic field to qualitatively explain the emergence of the chiral states,
and demonstrated that the exclusive energy region of the chiral state is a universal feature of 3D graphene systems.
We expect that the chiral states possibly make a significant contribution to the electronic transport 
in the randomly-connected 3D graphenes realized in the recent experiments.

\section*{Acknowledgments}

M. K. acknowledges support of JSPS KAKENHI Grant Numbers
JP17K05496, JP25107005 and JP25107001.

\bibliography{kiryu_3DQHE}

\end{document}